\documentclass[prb,aps,twocolumn,superscriptaddress]{revtex4-1}

\usepackage{times}
\usepackage{graphicx}
\usepackage{float}
\usepackage{latexsym,amsmath,amssymb,bm,euscript}
\usepackage{color}
\usepackage{epstopdf}
\usepackage[colorlinks=true,linkcolor=blue,citecolor=blue]{hyperref}
\usepackage{type1cm}
\usepackage[normalem]{ulem}

\begin{document}
\title{Ultrafast Laser Driven  Many-Body Dynamics and Kondo Coherence Collapse}

\author{W. Zhu}
\affiliation{Theoretical Division \& CNLS, Los Alamos National Laboratory, Los Alamos, New Mexico 87545, USA}
\affiliation{School of Science, Westlake University, Hangzhou, 310024, China and \\
	Institute of Natural Science, Westlake Institute of Advanced Study, Hangzhou, 310024, China}

\author{Benedikt Fauseweh}
\affiliation{Theoretical Division \& CNLS, Los Alamos National Laboratory, Los Alamos, New Mexico 87545, USA}

\author{Alexis Chacon}
\affiliation{Theoretical Division \& CNLS, Los Alamos National Laboratory, Los Alamos, New Mexico 87545, USA}

\author{Jian-Xin Zhu}
\affiliation{Theoretical Division \& CNLS, Los Alamos National Laboratory, Los Alamos, New Mexico 87545, USA}
\affiliation{Center for Integrated Nanotechnologies, Los Alamos National Laboratory, Los Alamos, New Mexico 87545, USA}

\begin{abstract}
Ultrafast laser pulse has provided a systematic way to inspect the dynamics of electrons in condensed matter systems.
In this paper, by means of time-dependent density matrix renormalization group, 
we study an ultrafast laser driven Kondo lattice model,
in which conduction electrons are strongly coupled with magnetically local moments.
The single-particle spectral function due to strong correlation effects and photon emission in the non-equilibrium states under laser driving are calculated.
We find  laser field excited collective doublon-hole pairs and an associated transient melting of Kondo coherence phase, signifying the collapse of Kondo energy gap. 
Moreover, we show that the photon emission, induced by a strong laser field, exhibits a different intensity characteristics than in the equilibrium Kondo insulator,
which could be explained by the Kondo collapse and related suppression of both intra-band and inter-band contribution in Kondo melting liquid. 
These theoretical insight is accessible with time- and angle-resolved photoemission spectroscopy and high-harmonic generation spectroscopy, and
will stimulate the investigation of nonequilibrium dynamics and nonlinear phenomenon in heavy fermion systems. 
\end{abstract}


\maketitle

\section{Introduction}
Interaction of intense ultrafast lasers with solids
 produces  extremely non-linear optical and electronic responses.~\cite{Krausz2000,Krausz2009,Krausz2017}
In particular, the laser-pulse driven electron dynamics allows to track real-time evolution of charge excitations,
thus making all-optical band structure reconstruction possible.~\cite{ZXShen2008,ZXShen2013,Vampa2015b}
Ultrafast pump-probe experiments have been successfully carried out on  elemental metals and semiconductors
\cite{Ghimire2011,Schubert2014,TTLuu2015,Ndabashimiye2016,Langer2016,HLiu2017,Tanaka2017,TTLiu2018,Reis2018,Hohenleutner2015}
and very recently in topological materials,~\cite{Bauer2018,Silva2018,Chacon2018}
where the electronic dynamics is usually described by a noninteracting electron theory.
It is now also believed that the technique can provide insight into collective behaviors 
in strongly correlated electronic systems.~\cite{Okamoto2007,Wall2011,MLiu2012,Mayer2015}
It leads to the exploration including the photoinduced high-temperature superconductivity,~\cite{DFausti:2011}
the long-lived photoinduced charge in the second-harmonic-generation signal from a transition-metal oxide heterostructure,~\cite{YMSheu:2013}
and the transient dehybridization in a $f$-electron heavy-fermion system~\cite{DLeuenberger:2018},
ultrafast demagnetization of magnetic materials involving	transition metals and rare earths \cite{Bigot1996,Jean2009,Marko2011,Luca2002,Rasing2010,Illg2011,Battiato2010,Carpene2008}, just to name a few.
In such strongly correlated systems, 
complex correlations is able to drive instabilities with macroscopic impacts like phase transitions and emergence of novel orders,
which makes the problem much more intricate. 

An ongoing quest of ultrafast spectroscopy is to investigate elementary excitations and their scattering processes, 
and novel phases  in strongly correlated quantum  materials driven out of equilibrium. 
The exciting progresses in this field of research have been made possible 
by the advances in spectroscopy probes like high-resolution time-resolved angle-resolved photoemission spectroscopy (tr-ARPES)~\cite{ZXShen2008,ZXShen2013,Vampa2015b}
and nonlinear optical high-harmonic generation 
(HHG).~\cite{Ghimire2011,Langer2016,Ndabashimiye2016,Tanaka2017,TTLuu2015,Schubert2014,HLiu2017,TTLiu2018,Reis2018}
In contrast to the well-established understanding in non-interacting electron systems,~\cite{Corkum1993,Maciej1994}
it is much more challenging to understand the underlying microscopic processes~\cite{Hughes1998,Tritschler2003,Golde2008,Ghimire2012,Kemper2013,Vampa2014,Vampa2015,Vampa2017,Tanaka2016,Rubio2017}
for HHG in strongly correlated materials,
where electron, spin, and orbital degrees of freedom are fundamentally entangled.
Though several attempts were made,~\cite{Oka2012,Ivanov2018,Yuta2018,Fahnle2018}
it is far from clear the role of many-body strong correlation effects and how they influence the ultrafast electron dynamics.
Therefore, it is highly desirable to investigate how the coupling between different degrees of freedom 
would influence ultrafast electron dynamics in strongly correlated systems, 
and particularly identify the roles of the driving field parameters (e.g. intensity, periodicity) 
for the novel quantum phase transition out of equilibrium.

Heavy-fermion compounds containing rare-earth or actinide elements are 
a canonical condensed matter system with strong correlations. 
In these systems,  the interplay between  strong on-site Coulomb repulsion on localized $f$-orbital electrons and their hybridization with conduction electrons 
in other $s$-, $p$-, $d$-orbitals~\cite{Hewson1997,Coleman2015} gives rise to emergent phenomena such as unconventional superconductivity and local quantum critical point.~\cite{Qimiao2008}. 
Despite of some attempts on ultrafast dynamics in the heavy-fermion compounds \cite{NLWang2016,DLeuenberger:2018}, 
to our best knowledge, a thorough theoretical study is still lacking.

To fill in this gap, in this paper, we study ultrafast laser-pulse driven many-body dynamics in the Kondo lattice model (KLM),
which is a prototypical model describing the essential  heavy-fermion physics. 
Using time-dependent density-matrix renormalization group algorithm (TD-DMRG),~\cite{White1992,White2004,Vidal2004,Feiguin2005} we demonstrate the complex many-body electron dynamics and transient phase transition stemming from the Kondo coherence collapse,
which are recorded by real-time evolution of doublon-hole pair population and effective quasiparticle band dispersion.
We identify the re-collision between the excited doublon and its associated hole, which generates high-harmonic emission spectra. We also address field intensity dependence of the cutoff of emitted photon energy, field dependence of spectrum intensity and time-resolved emission spectrum.
Our findings provide useful informations for understanding the hybridization dynamics 
and non-linear behavior in heavy-fermion systems using intense ultrafast pump and probe techniques.

\section{Model and Method}
We consider a half-filled KLM~\cite{Ueda1993,Yu1993,Ueda1997,Doniach1977} 
in the presence of an intense laser field
\begin{equation}\label{eq:ham}
\hat H=  \sum_{ i,\sigma}^{L-1}[ J_0 e^{-i\Phi(t)} c^{\dagger}_{i\sigma} c_{i+1,\sigma} +h.c.] + J_K \sum_i \mathbf S_i \cdot \mathbf s_i\;.
\end{equation}
Here $c^\dagger_{i\sigma}$ denotes the creation operator of a conduction electron with spin $\sigma=\uparrow,\downarrow$,
and $\mathbf s_i$ ($\mathbf S_i$) is the effective spin of conduction electron (localized spin moment). 
The conduction electrons interact with local spin-$\frac{1}{2}$ moments $\mathbf S_i$ via a Kondo coupling $J_K$. 
The hopping parameter of electron bath is set to be $J_0 =250$ meV$\approx 0.00919$a.u., so that the band width in the conducting limit is
$W_0 = 4J_0 = 1$ eV $\approx 0.0367$a.u.. 
In this paper, we will frequently use atomic unit and natural unit to serve for the readers from different backgrounds.
In Eq.~(\ref{eq:ham}), the laser electric field $E(t)=-dA(t)/dt$ is incorporated through 
the time-dependent Peierls phase \cite{JXZhu_Book} $ea_0 E(t)=-d\Phi(t)/dt$ where $a_0$ is the lattice constant and $e$ is electric charge unit. 
The Peierls substitution is widely used in simulating the coupling of the coupling between electrons and laser electromagnetic field physics.~\cite{Devereaux2012,Devereaux2013,Devereaux2014,Devereaux2018,Devereaux2019,Werner2017}
To put our model in a solid footing, we have given a detailed derivation in Appendix A for the validity of this treatment and demonstrated the basis in Eq.~(1) is time-independent.
Throughout the work, we choose $a_0\approx 7.56$a.u.$\approx 4 \AA$, which is close to that in newly synthesized Kondo materials CeCo$_2$Ga$_8$.~\cite{LWang2017}
The field vector potential is $A(t)=A_0 \sin^2(\frac{\omega_0 t}{2N_0}) \sin(\omega_0 t)$,  
which encloses a total duration of $N_0$ optical cycles in the time domain $t\in [-N_0\frac{2\pi}{\omega_0},0]$.
All parameters of the pulse are well within experimental reach, i.e. typical frequency $\omega_0\in [0.005,0.014]$a.u. (equivalent to $[30,90]$THz)
and the laser peak amplitude $E_0=A_0\omega_0\in[0.001,0.008]$a.u.(equivalent to $[5, 40]$ MV$\cdot$cm$^{-1}$).

The unitary time-evolution is simulated by TD-DMRG algorithm.~\cite{White1992,Vidal2004,White2004,Feiguin2005}
After obtaining the steady ground state $|\Psi_{-\infty} \rangle=|\Psi(t=-\infty) \rangle$,
we let the system evolve according to the time-dependent $H(t)$ and compute the wave function
$|\Psi(t)\rangle=U(t,-\infty) |\Psi(t=-\infty) \rangle$ and time-evolution operator $U(t,t_0)= e^{-i\frac{e}{\hbar} \int^t_{t_0} dt' H(t')}$.
The time step for calculation is $\Delta t=0.25 a.u. \approx 6$ attosecond.
In the calculations, we implement the second-order Trotter-expansion, and use up to $M=1024$ bond dimension
(the resulting discard error is at the level of $10^{-5}$).
We choose the system size of  $L=24,36,48$ and find the qualitatively same picture is reached. The results shown below are based on $L=48$ size.

The single-particle spectral function can be defined by
the probability for an electron to emit into an unoccupied state (or absorb from occupied state) in response to a probe pulse starting at $t=0$, before which the pump pulse has been completed:
\begin{equation}
A(i,j,t_{\mathrm{f}})= \langle \Psi_{0}|\{c^{\dagger}_{i,\sigma}(t_{\mathrm{f}}), c_{j,\sigma}(0) \} | \Psi_{0}\rangle\;,
\end{equation}
where $c^{\dagger}_{i,\sigma}(t)= e^{iHt} c^{\dagger}_{i,\sigma} e^{-iHt}$ 
and $t_{\mathrm{f}}$ is the final time for measurement.
The momentum-frequency resolved spectral function is obtained by Fourier transform:
$A(k,\omega) = \frac{1}{L_0} \sum^{L_0}_{r=1} e^{i k r} \int^{t_{\mathrm{f}}}_{0} dt e^{i\omega t-\eta t} A(0,r,t)$.
We set the smearing energy $\eta=0.2$ (in units of electron hopping energy $J_0$).
To reach good resolution in time domain Fourier transform, we choose $|t_{\mathrm{f}}| > 1000$ a.u.
We also choose a segment enclosing $L_0=12$ sites in the middle of a finite chain to perform the Fourier transform in spatial space.~\cite{note1}

The pulse induced current operator is 
\begin{equation}
\hat J(t)=-iea_0 J_0 \sum_{i\sigma} [e^{-i\Phi(t)}c^\dagger_{i,\sigma}c_{i+1,\sigma}-h.c.]\;.
\end{equation}
The harmonic emission is obtained by the square of Fourier transform of  dipole acceleration\cite{Ivanov2018,SHLin} $|\mathcal{FT}[d\langle J(t)\rangle/dt]|^2$.
Time-resolved harmonic emission is calculated by performing the Gabor transform (window Fourier transform) with the sliding 
window $e^{-(t-t^{\mathrm{target}})/\sigma^2}$ by setting $\sigma=1/(3\omega_0)$.


\begin{figure}[t]
  \includegraphics[width=0.97\linewidth]{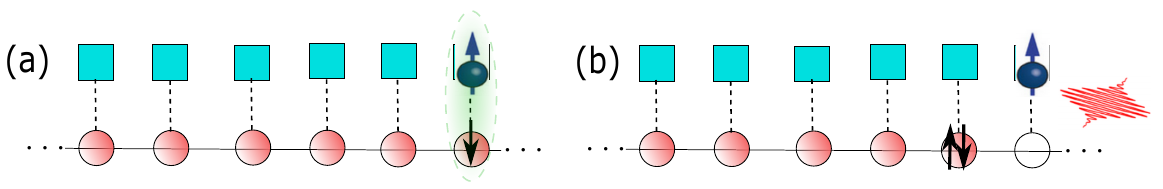}
  \includegraphics[width=0.99\linewidth]{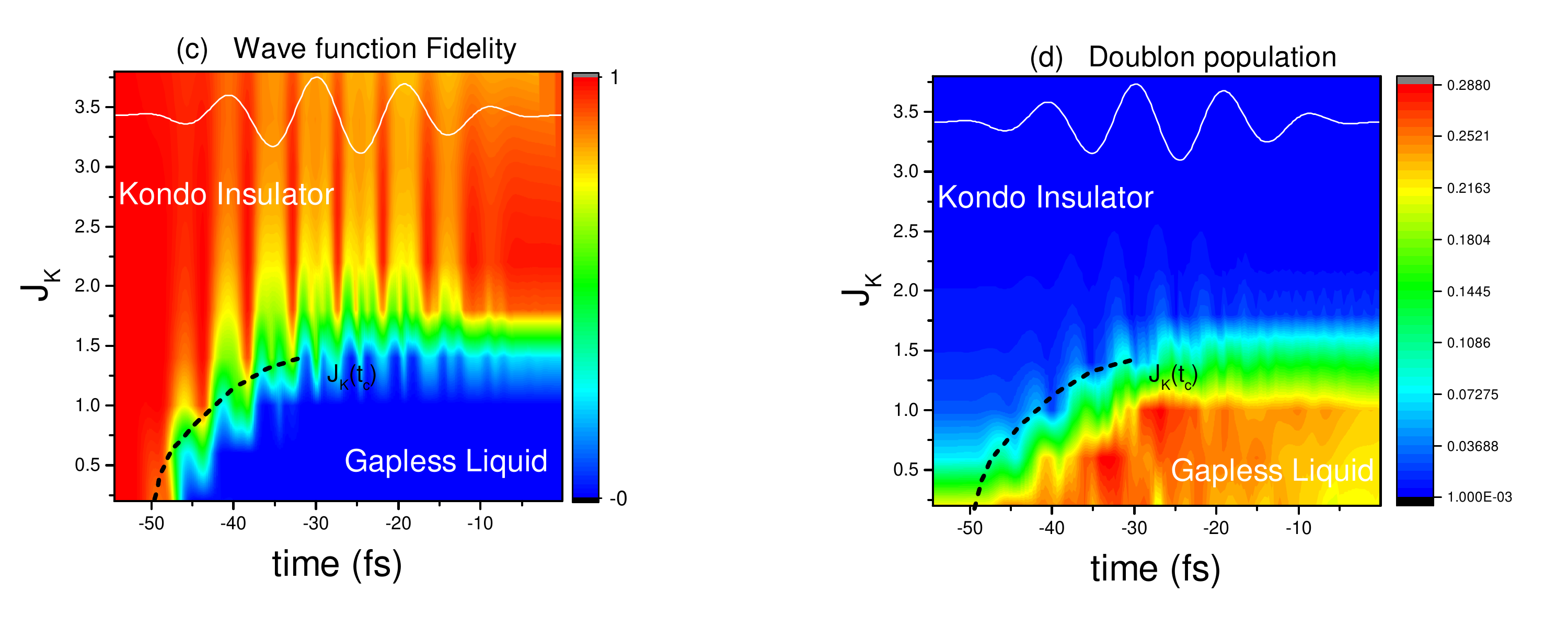}
  \caption{\textbf{Laser pulse induced breakdown of the Kondo insulator.}
  (a) Cartoon picture for the equilibrium ground state of Kondo insulator with Kondo singlet forming on each site. The blue square (red circle) represents local moment (electron) degree of freedom.  (b) Laser field excites double-occupancy and hole states, which leads to the collapse of Kondo singlet pairing.
  (c) Time-resolved wave-function overlap with the initial state $G(t)=|\langle \Psi(t=-\infty)|\Psi(t) \rangle|^2$,
  where $\Psi(t)$ is the time-evolution wave function.
  (d) Time-resolved averaged double occupancy $d_{j=L/2}= \langle \Psi(t)|n_{j,\uparrow} n_{j,\downarrow}| \Psi(t)\rangle$.
  We set the pulse frequency $\omega_0=0.014a.u.\approx90$THz, peak strength $E_0=0.005a.u.\approx 25$MV cm$^{-1}$ and a $5-$cycle $\sin^2$ envelope (white solid line).
  The black dashed line shows the time when the threshold field strength $E_{\mathrm{th}}$ is reached.
  } \label{fig:population}
\end{figure}

\section{Breakdown of Kondo insulator}
We start by discussing the salient features induced by applying a laser pulse. Generally, in equilibrium case the ground state of the KLM is a product state of Kondo singlets (see Fig. \ref{fig:population}(a)), where each site hosts a spin-singlet state forming by one local spin moment and one conduction electron. Applying
the intense laser pulse can excite the charge carriers, resulting in a hole on one lattice site (a holon) and a neighbouring site with two electrons (a doublon) (see Fig.~\ref{fig:population}(b)).
The creation of doublon and holon carriers will destroy the Kondo singlets.
In this context, the laser field is expected to induce the collapse of the Kondo insulator.

We show time-resolved laser-induced breakdown of the Kondo insulator, via the time-evolution of wave function fidelity
$G(t)=|\langle \Psi(t=0)|\Psi(t)\rangle|^2$ (Fig.~\ref{fig:population}(c))
and the averaged density of doublon,
$d_{j=L/2}= \langle \Psi(t)|n_{j,\uparrow} n_{j,\downarrow}| \Psi(t)\rangle$ (Fig. \ref{fig:population}(d)).~\cite{note2} 
First of all, we identify the quantum phase transition depending on the Kondo coupling strength $J_K$. 
In small $J_K$ regime,
it is found that the wave function fidelity $G(t)$ drops to nearly zero within several laser cycles. Simultaneously, the population of doublons rises and saturates to $d_j^{\mathrm{sat}}\sim 0.25$. Both results demonstrate the destruction of the Kondo insulator during the pulse.
On the contrary, for large $J_K$ regime,
wave function fidelity $G(t)$ is close to one (despite some regular oscillations) and $d_j(t)$ is always pinning at zero,
indicating the failure of exciting doublon-holon pairs.

\begin{figure*}[t]
	\includegraphics[width=0.75\textwidth]{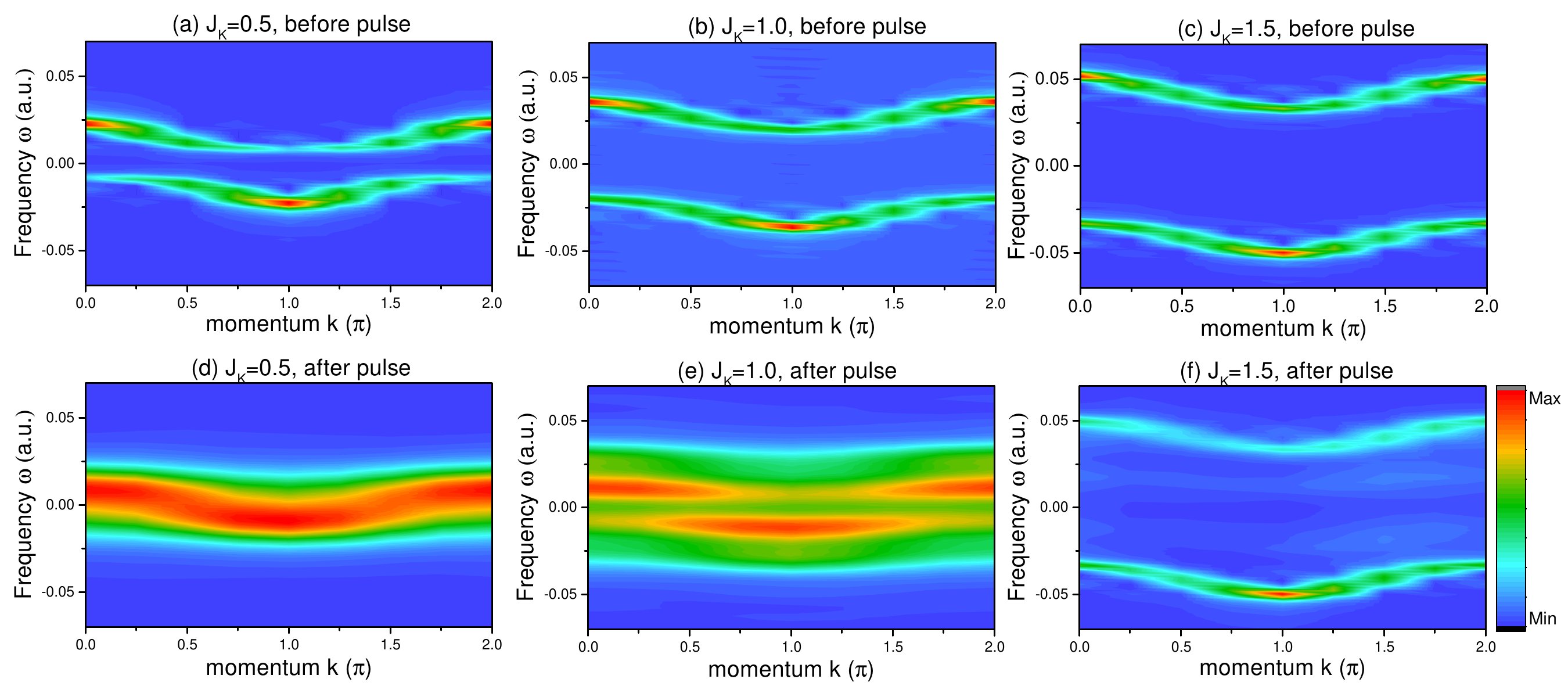}
	\caption{\textbf{Energy-momentum resolved spectra.}
		Contour plots of spectral function $A(k,\omega)$ of the conduction electron (top) without a driven pulse field
		and (bottom)  with a driven pulse field, for various Kondo couplings: (a,d) $J_K=0.5$ eV$\approx 0.0183$a.u., (b,e) $J_K=1.0$ eV$\approx 0.0367$a.u. and (c,f) $J_K=1.5$ eV$\approx 0.0551$a.u..
		The laser has peak strength $E_0=0.005$ a.u., frequency $\omega_0=0.014$ a.u. and a  5-cycle sine-square envelope.
	} \label{fig:Akw}
\end{figure*}

To understand the above observations, we perform an adiabatic perturbative analysis based on Landau-Zener theory.~\cite{Landau1932,Zener1932}
In the quantum tunneling regime (non-resonance limit $\omega_0 \ll \Delta$), we obtain a threshold field $E_{\mathrm{th}}$ in the dc-limit  (see Appendix B):
$ E_{\mathrm{th}}\approx \frac{\Delta^2}{4v}$,
where $\Delta$ is the Kondo excitation gap of electrons which is dependent on $J_K$ and $v$ is group velocity.~\cite{Ueda1993,Yu1993,Ueda1997,Oka2008}
Physically,  $E_{\mathrm{th}}$ characterizes that quantum tunneling process dominates the creation of doublon-holon pairs, when the energy drop in length scale equivalent to electron correlation length $\xi$ becomes comparable to the excitation gap $\xi E_{\mathrm{th}}\sim\Delta$.~\cite{Oka2005,Oka2008}
Then the transition is expected at when the condition of $E(t_c)=E_{\mathrm{th}}$ can be satisfied, where $t_c$ is the first time that electric field $E(t)$ exceeds the threshold field $E_{\mathrm{th}}$. In Fig.~\ref{fig:population}(c-d),
we plot the dependence of $J_K$ on transition time $t_c$ (black dashed line), estimated from Landau-Zener theory. As we see, if $E(t)$ exceeds the threshold value $E_{\mathrm{th}}$,
the driving laser field can induce a quantum phase transition and the Kondo insulator breaks down.
Otherwise, if the laser field is not strong enough ($E(t)<E_{\mathrm{th}}$), the Kondo insulator is stable against to the laser pulse.
Importantly,
the emergence of threshold field $E_{\mathrm{th}}$ and its dependence on excitation gap $\Delta$ indicate that,
the Kondo insulator with larger Kondo coupling strength is more robust against  a laser pulse field,
consistent with our observations in Fig.~\ref{fig:population}.


\section{Spectral function}
To strengthen the evidence leading to our previous assignment,
we  show the effects of laser pulses on spectral functions in Fig. \ref{fig:Akw} in the separated panels.
For reference, we first show the cases before arrival of pulse field (as illustrated in Fig. \ref{fig:Akw}(a-c)),
which probe the equilibrium electronic structures.
Clearly, there exists a spectral gap in equilibrium spectral function for all finite $J_K$, which manifests the hybridization induced Kondo insulator.~\cite{Ueda1997}
Importantly, the results after applying the pulse field reveal a dichotomy  as one tunes $J_K$.
For small $J_K$ region (Fig.~\ref{fig:Akw}(d)),
after pumping, it is found that the spectral gap is completely smeared out, and the overall spectral weight shows a considerable shift toward Fermi energy (setting at $E_F=0$).
The closing of the Kondo energy gap through transient recovery of the gapless electronic dispersion
indicates the melting of Kondo insulating state.
In addition, the effective band dispersion is significantly broadened after pumping, which can be attributed to the multi-scattering processes by localized spins.
In large $J_K$ region (Fig.~\ref{fig:Akw}(f)), the Kondo coherent bands become stiff, despite some photo-excited states appear in the Kondo gap and partial spectral weight transfers to the low-frequency region.
With the increase of $J_K$, we identify a smooth crossover  from laser-driven gapless liquid $J_K\lesssim 1.25$ eV to the Kondo insulator phase $J_K\gtrsim 1.25$ eV (Fig. \ref{fig:Akw}(d-f)).
Again, these observations demonstrate that the Kondo insulator with smaller gap is more susceptible to the photo-excitation.

Figure~\ref{fig:Akw} outlines the key features that allow us to extract the effective quasi-particle dispersion, which can be directly mapped out by the time and angular-resolved photo-emission spectrum (tr-ARPES) measurements.~\cite{DLeuenberger:2018}
In addition, the effective quasi-particle dispersion is insightful to understand the non-linear optical response, as we will show below.


\begin{figure*}[t]
  \includegraphics[width=0.85\textwidth]{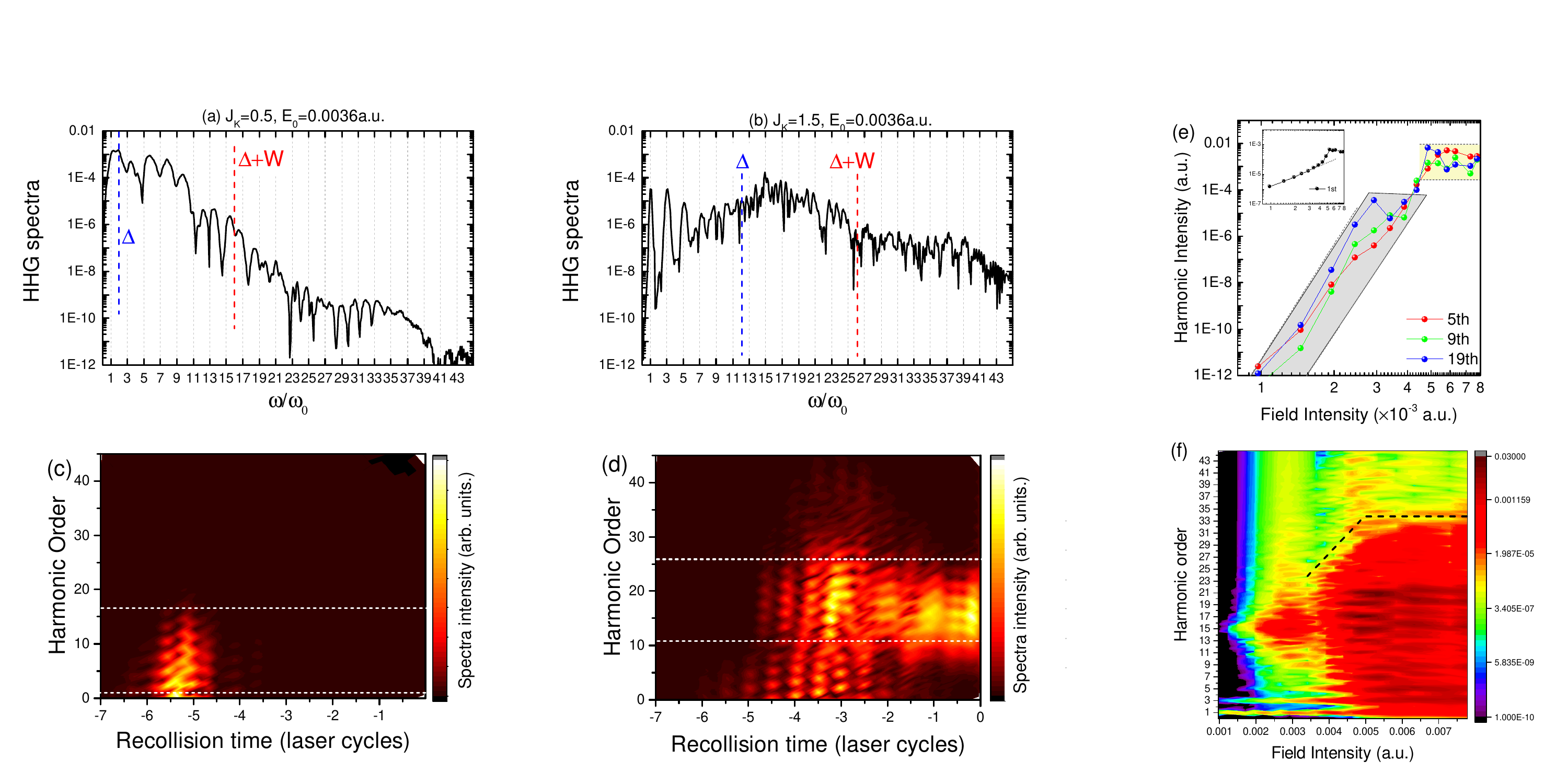}
  \caption{\textbf{High-harmonic spectroscopy of laser-pulse driven  Kondo lattice model.}
  Harmonic spectrum for (a) $J_K=0.5$ eV$\approx 0.0183$a.u. and (b) $J_K=1.5$ eV$\approx 0.0551$a.u.. The vertical dashed lines show energy scales of $\Delta$ (blue) and $\Delta+W$ (red), where charge excitation gap $\Delta$ and effective band width $W$ are determined by spectral functions in Fig.~\ref{fig:Akw}(a,c).
  Gabor profiles of harmonic order as a function of time for (c) $J_K=0.5$ eV$\approx 0.0183$a.u. and (d) $J_K=1.5$ eV $\approx 0.0551$a.u.. 
  The white dashed line in (c) and (d) mark the energy gap $\Delta$ and $\Delta+W$, related to the dashed line in (a) and (b). 
  (e) The intensity of the fifth (red), ninth (green) and nineteenth (blue) harmonic order dependence on the field strength.
  (f) Contour plot of harmonic spectra and various field intensity. 
  The laser field has the pulse frequency $\omega_0=0.005$ a.u. (30 THz), peak strength $E_0=0.0036$ a.u. 
  ($\sim20$ MV cm$^{-1}$) and a 7-cycle sine-square envelope. 
  } \label{fig:HHG}
\end{figure*}

\section{High-harmonic Spectroscopy}
The appearance of photo-excited charged carriers is expected to generate the non-linear optical responses.
Figure~\ref{fig:HHG} shows a typical analysis of characteristic high-harmonic emissions manifesting
 the charged excitations.
Generally, compared with the non-interacting case that shows well-defined odd harmonic structures,~\cite{Ghimire2012,Kemper2013,Vampa2014}
the harmonics spectra of the driven Kondo insulator is less regular.
Despite the noisy spectra, several universal behaviors can be identified.
In small $J_K$ regime (Fig.~\ref{fig:HHG}(a)) the hamonics are largely bounded by the effective band width $W$, below which the harmonic structures can be attributed to the intra-band current   in the effective single-band dispersion (see Fig. \ref{fig:Akw}(d)).
In large $J_K$ regime (Fig.~\ref{fig:HHG}(b)), increasing  Kondo hybridization suppresses low-order harmonics and generate high-order harmonics above the energy gap $\Delta$, where the high-harmonic emissions should be associated with the inter-band current in the effective two-band dispersion (see Fig.~\ref{fig:Akw}(f)).
In comparison with the HHG spectra of the Kondo melting phase (Fig.~\ref{fig:HHG}(a)), one notable difference is that  the spectra of Kondo insulator (Fig.~\ref{fig:HHG}(b)) is much sharper. In Fig.~\ref{fig:HHG}(b), one can see well-resolved  odd-harmonic signal peaks below energy gap $\Delta$.
This can be explained by much long (short) relaxation time of excitations in the Kondo insulator (Kondo melting phase), which is again consistent with energy-momentum resolved spectrum function as shown in Fig.~\ref{fig:Akw}: The effective band is significantly broadened in the Kondo melting phase but not for the  hard Kondo insulator.

To further understand the dynamics of harmonics generation in more detail, we proceed by analyzing the time profiles of emission spectra (Fig.~\ref{fig:HHG}(c-d)), allowing for the investigation of the harmonic emission with sub-cycle temporal resolution.
We find several notable differences. Firstly, for small $J_K$ (Fig.~\ref{fig:HHG}(c)),
the photon emission occurs at the first and second circles of driving field, synchronizing with
the dynamical breakdown of the Kondo insulator (see Fig. \ref{fig:population}).
In contrast, for large $J_K$ (Fig.~\ref{fig:HHG}(d)), the photon emission occurs around the peak of driving field.
Since the Bloch oscillation is strongly suppressed, the strong photon emission should result from inter-band contributions. 
Secondly, comparing Fig.~\ref{fig:HHG}(c) and (d) in detail, we identify signals of 
recollision trajectories of doublon-holon pairs in large $J_K$ region.
These trajectories can be understood by the interband recollisions, i.e.
doublons can recollide with the hole at a specific time. 
Upon recombination of the doublon and  holon, energy difference is transferred to high harmonic photons.
This result also suggests that emission is generated by one-photon transitions back the original ground state through single double-hole recombination 
(see Appendix C).
These differences in time-profile analysis again confirm that, after applying laser field, the small $J_K$ region is gapless liquid described by an effective single-band dispersion, while the large $J_K$ region corresponds to a Kondo insulator with robust hybridization gap.

As last, we present the emission spectra intensity and cutoff dependence on the field strength. As functions of the strength of driving field, Fig.~\ref{fig:HHG}(e) shows that harmonic intensities approximately
follow a $E_0^p$ scaling ($p\approx 11$) regardless of harmonic orders in weak field regime (light blue shade), whereas it is close to saturation in high field regime (light yellow shade). This dependence considerably deviates from the atomic limit~\cite{Corkum1993,Maciej1994} indicating a strongly non-perturbative regime of light-matter interactions.
Moreover, in  Fig. \ref{fig:HHG}(f), in a wide range of field strengths, the harmonic spectra display well-resolved plateaus, followed by a cutoff. For strong field strength, the cutoff is field-independent (black dashed line).

The laser pulse parameters reported here are experimentally accessible.
For instance, the time-resolved high harmonic emissions with $\sim1$ fs accuracy are experimentally 
achievable.~\cite{Hohenleutner2015}
This is sufficient to distinguish the Kondo insulator from the Kondo melting phase by using the novel nonlinear HHG features obtained in this work.

\section{Discussion and Summary}
In summary, we have presented a thorough study of the transient collapse of the Kondo insulator and documentation of the melting Kondo energy gap, 
by applying an ultrafast laser field.
We were able to track electric and optical response in real time.
Importantly, knowledge of the trajectory of  recolliding doublon-holon pair
links photon energy to the effective band dispersion, is close agreement with the energy-momentum resolved electronic structures.
Thus, our results show that, in the heavy-fermion systems the breakdown of Kondo insulator (collapse of local Kondo singlet) can be 
	inspected by the ultrafast dynamics such as HHG.

Although, in this paper, due to the computational limit, the theoretical model is greatly simplified (e.g. we only consider a single conduction band, and 
	starting equilibrium phase is a Kondo insulator (at the half-filling of conduction electrons)),
we make an essential step to study ultrafast dynamics in heavy-fermion systems. 
Our findings have implications well beyond the specific model,
which may impact the non-linear dynamics of charge carriers in complex materials such as heavy-fermion systems in which several degrees of freedom play a critical role.
On the one hand, the fundamental many-body dynamics obtained here also sheds lights on high-dimensional strongly-correlated 
$f$-electron systems~\cite{Fiebig2018,Pal2018} (also see discussion in Appendix. D).
On the other hand,  we believe the results reported in this work will stimulate experimental realization of
one-dimensional heavy-fermion systems (e.g.  CeCo$_2$Ga$_8$~\cite{LWang2017}) for the study of non-equilibrium 
many-body dynamics.
In future, it is deserved to further study ultrafast dynamics in 
heavy-fermion systems, by considering more realistic experimental conditions.

\textit{Acknowledgments.---}
We thanks  Chen-Yen Lai  for fruitful discussion.
This work was carried out under the auspices of the National Nuclear Security Administration of the U.S. Department of Energy at Los Alamos National Laboratory (LANL). 
It was supported by the LANL LDRD Program.

\bibliographystyle{apsrev}

\begin{thebibliography}{50}
\bibitem{note1} 
The sites in the middle of a finite chain automatically recovers the translational symmetry since $A(i,j,t,t_i)\approx A(i-j,0,t,t_i)$. Therefore, we can glue the middle segment into a ring and approximately define the momentum quantum number.
\bibitem{note2}The key feature of Kondo coherence is the formation of local entangled singlet. The breakdown of this local Kondo singlet 
will lead to a quantum phase transition and breakdown of Kondo insulator. 
This is the general reason to inspect the local terms like doublon occupancy $d_j$. 
$d_{j=L/2}$ is a typical choice, without loss of generality, to show the physics here. 
We have confirmed that the main features discussed in this work are the same on different local sites.
\end{thebibliography}


\clearpage
\begin{widetext}
\noindent {\bf Supplementary Information}
\appendix

\section{Peierls substitution derivation}

In this section we show how the Peierls form for time-dependent fields $\mathbf{A}(t)$ can be derived.
We focus on the conduction orbital and write down the Hamiltonian in terms of the second quantized field operators,

\begin{equation}
\label{eq.FieldOperator}
\begin{aligned}
H(t) = \int \mathrm{d}\mathbf{r} {\Psi}^\dagger(\mathbf{r}) \left[ \frac{\left(-i \hbar \nabla - e \mathbf{A}(t)\right)^2 }{2m} + V(\mathbf{r})  \right] {\Psi}(\mathbf{r}),
\end{aligned}
\end{equation}

where the explicit time dependence comes from a time-dependent vector potential whereas the second quantized field operator is time independent. Here we have also suppressed the spin index for the simplicity of the discussion. Now we express the field operator in terms of the atomic orbital basis,

\begin{equation}
\label{eq.Wannier}
\begin{aligned}
{\Psi}(\mathbf{r}) = \sum\limits_l \phi(\mathbf{r}-\mathbf{R}_l ) c_l
\end{aligned}
\end{equation}

Here $ \phi(\mathbf{r}-\mathbf{R}_l )$  is the atomic-like orbital, e.g., Wannier orbitals, at the atomic site $l$  and $c_l$   is the annihilation operator of one electron at site. At this stage, there is no time dependence on the field operator ${\Psi}(\mathbf{r})$, and there is no time dependence on the $\phi(\mathbf{r}-\mathbf{R}_l )$ and the operator $c_l$ either. 
We substitute Eq.\ \eqref{eq.Wannier} into Eq.\ \eqref{eq.FieldOperator}, and obtain 
\begin{equation}
\begin{aligned}
H(t) = \sum\limits_{l l'} J_{l l'} (t) c_l^\dagger c_l^{\phantom{\dagger}}
\end{aligned}
\end{equation}
where the hopping integral is given as 
\begin{equation}
\label{Eq.HoppingIntegral}
\begin{aligned}
J_{l l'} (t) = \int \mathrm{d} \mathbf{r} \phi^*(\mathbf{r}-\mathbf{R}_l) \left[ \frac{(-i \hbar \nabla - e A (t) )^2}{2m} + V(r) \right] \phi(\mathbf{r}-\mathbf{R}_{l'}).
\end{aligned}
\end{equation}
Given a general Hamiltonian $H(\mathbf{r},\mathbf{p})$ as function of position and momentum, the minimal coupling transformation can be written as a gauge transformation,
\begin{align}
\exp \left( -i e \int^{\mathbf{r}} \mathbf{A}(t) \mathrm{d} \mathbf{s} \right) H(\mathbf{r},\mathbf{p}) \exp \left(i e \int^{\mathbf{r}} \mathbf{A}(t) \mathrm{d}  \mathbf{s} \right) = H(\mathbf{r}, \mathbf{p} - e\mathbf{A}(t)).
\end{align}
Under the assumption that the Wannier orbitals are strongly localized and within the long wave length approximation, i.e. the $\mathbf{A}$ field is spatially uniform, we can approximate the Hopping integral \eqref{Eq.HoppingIntegral} as
\begin{align}
J_{l l'} &= \int \mathrm{d} \mathbf{r} \phi^*(\mathbf{r}-\mathbf{R}_l) \exp \left( -i e \int^{\mathbf{r}} \mathbf{A}(t) \mathrm{d}\mathbf{s} \right) \left[ \frac{(-i \hbar \nabla )^2}{2m} + V(r) \right] \exp \left(i e \int^{\mathbf{r}} \mathbf{A}(t) \mathrm{d}\mathbf{s} \right) \phi(\mathbf{r}-\mathbf{R}_{l'}) \\
&\approx e^{ i \int_{\mathbf{R}_l}^{\mathbf{R}_{l'}} e\mathbf{A}(t) \mathrm{d}\mathbf{r'}}  \int \mathrm{d} \mathbf{r} \phi^*(\mathbf{r}-\mathbf{R}_l)  \left[ \frac{(-i \hbar \nabla )^2}{2m} + V(r) \right]  \phi(\mathbf{r}-\mathbf{R}_{l'}) \\
&= e^{ i \int_{\mathbf{R}_l}^{\mathbf{R}_{l'}} e\mathbf{A}(t) \mathrm{d} \mathbf{r'}}  J_{l l'}^0
\end{align}
which is the Peierls form of the hopping.

\section{The threshold field $E_{\mathrm{th}}$}
\label{app:Eth}
In this section, we will estimate the threshold field $E_{\mathrm{th}}$ in the DC-limit based on the effective hybridization bands in Kondo lattice model in the framework of the Landau-Zener's method.
Fig.~\ref{sfig:Eth}(a) shows dispersion relation of conduction electrons (brown) and a flat band (blue).
Hybridization leads to form an indirect energy (hybridization) gap due to the coherent Kondo screening of the local moments by the sea of conduction electrons. In case of Kondo insulators the Fermi level is located in the hybridization gap, thus the band below Fermi level is full occupied. Note that, the cartoon picture in Fig.~\ref{sfig:Eth}(a-b) has been confirmed by our DMRG calculations (see Fig.~2 in main text). The resulting band dispersion contains two bands (green in Fig.~\ref{sfig:Eth}(a)), which will be the starting point for the Landau-Zener's analysis.~\cite{Landau1932,Zener1932}

\begin{figure}[b]
	\includegraphics[width=0.47\textwidth]{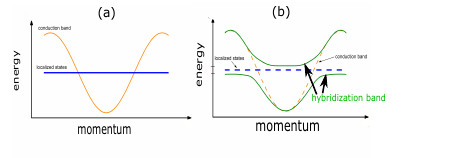}
	\includegraphics[width=0.17\textwidth]{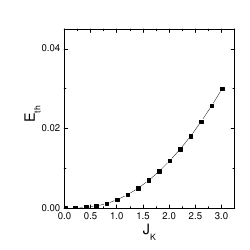}
	\includegraphics[width=0.165\textwidth]{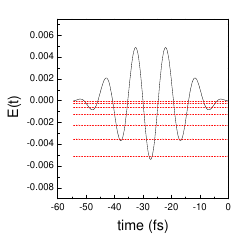}
	\caption{(a) Dispersion relation of conduction band (brown) and a localized flat band (blue). (b) Hybridization induced effective band structures (green).
		(c) The dependence of estimated threshold field $E_{\mathrm th}$ on Kondo coupling strength $J_K$. The charge excitation gap is obtained on system size $L=48$ Kondo lattice model.
		(d) The time-profile of electric field $E(t)$ (black line),  where the laser field has peak strength $E_0=0.006$ a.u., frequency $\omega_0=0.014$ a.u. and a 5-cycle $\sin^2$ envelope.
		The estimated threshold field $E_{\mathrm th}$ (red dashed line)  for various Kondo couplings (from top to bottom) $J_K=0.2,0.4,0.6,0.8,1.0,1.2$. The first crossing between $E(t)$ and $E_{\mathrm th}$ gives the transition time $t_c$.
	} \label{sfig:Eth}
\end{figure}

Following adiabatic perturbation theory, we have
\begin{eqnarray}
\hat H(\Phi)|0;\Phi\rangle&=&E_0(\Phi)|0;\Phi\rangle,\\
\hat H(\Phi)|1;\Phi\rangle_{dh}&=&E_{dh}(1;\Phi)|1;\Phi\rangle_{dh},
\end{eqnarray}
where $|0;\Phi\rangle$ and $|1;\Phi\rangle_{dh}$ respectively denotes
the ground state without a doublon-hole pair and excited state with one doublon-hole pair.
By ignoring multiple pair states,
we only consider the  excitation channel with one doublon-hole pair.
To simplify the problem, we can study the tunneling process in a Hilbert space spanned by two states, thus
the problem reduces to solving the time-dependent Schr\"odinger
equation of the form
\begin{eqnarray}
|\Psi(t)\rangle=a(t)|0;\Phi(t)\rangle+b(t)|1;\Phi(t)\rangle_{dh}
\end{eqnarray}
and the initial condition is $a(0)=1,b(0)=0$.
Next we consider the dc electric field: $E(t)=E_0, d\Phi(t)/dt=E_0$. With the help of the Landau-Zener's theory,
the tunneling probability from the ground state to the excited state is
\begin{equation}
\mathcal{P}_p^{Landau-Zener}=|b(t\rightarrow \infty)|^2=\exp(-\pi \frac{E_{\mathrm th}}{E_0}),\,\,\,\,\,
E_{\mathrm th}= \frac{\Delta^2}{4v}.
\label{eq:LZ}
\end{equation}
Here, $\Delta$ is the charge excitation gap between conduction band and valence band, $v$ is the group velocity. The key message of Eq.~(\ref{eq:LZ}) is, the threshold field $E_{\mathrm th}$ quadratically depends on energy gap $\Delta$.

Next, we will numerically determine the excitation gap $\Delta$ in Kondo lattice model and estimate $E_{\mathrm th}$.
In Kondo lattice model, since there are spin and charge degrees of freedom, the typical energy scales include spin gap and charge gap.
In this work, we focus on the charge excitation gap, since the laser electric field
is directly coupled to electrons. The charge gap is defined by the difference of the lowest energy in the subspace
$N_e=L$ and $N_e=L+2$:
\begin{equation}
\Delta=E_0(N_e=L+2,S^z=0)-E_0(N_e=L,S^z=0).
\end{equation}
Owing to the hidden $SU(2)$ symmetry in the charge space, the energy
difference is the same as the charge excitation gap in the subspace of the fixed number of electrons $N_e=L$.~\cite{Ueda1993}
After we determine the charge gap, we can get the threshold field $E_{\mathrm th}$ using Eq.~(\ref{eq:LZ}). In Fig. \ref{sfig:Eth}(c), we show the dependence of $E_{\mathrm th}$ on Kondo coupling strength $J_K$ on the system size $L=48$.

Although Eq.~(\ref{eq:LZ}) is obtained at the dc-limit, we find it is helpful to understand the physics under the influence of time-dependent laser field. In Fig.~\ref{sfig:Eth}(d), we plot the time-profile of electric field $E(t)$ (black) and threshold value $E_{\mathrm th}$ for various $J_K$. Here, we reach the condition of $E(t_c)=E_{\mathrm th}(J_K)$, where $t_c$ is the first time that electric field exceeds the threshold field.
In general, the larger $J_K$, the longer critical time $t_c$ is expected, as shown in Fig.~\ref{sfig:Eth}(d). This basically gives us an understanding why the $J_K$ dependent laser-induced phase transition in the main text (Fig.~1).
The threshold value $E_{\mathrm th}$ shown in Fig.~1 (white dashed line) is estimated using Eq.~(\ref{eq:LZ}).




\section{Semi-classical analysis of recollision trajectory}
Under a single active electron approximation and a two-band model, usually the inter-band current determines the highest harmonics emitted by a laser driven semiconductor media.~\cite{Vampa2017} Hence, the electron-hole trajectories three steps classical approach could predict the harmonic emission along the plateau and cutoff.~\cite{Vampa2015,Vampa2017} Here, we assume the band structure of conduction band and valence band as
\begin{equation}
\epsilon_{\mathrm c}(q)= \Delta+v_{\mathrm c} \cos q,\,\,\,\,\, \epsilon_{\mathrm v}(q) = -\Delta+v_{\mathrm v} \cos q.
\end{equation}
The electron and hole wave vector relative to its value at $t=0$,
\begin{eqnarray}
q(t)&=& -\frac{e}{\hbar} \int^t_0 E(t') dt' = A(t)+ q(0), 
\end{eqnarray}
Here, $E(t)$ is the electric field of the laser and $A(t)$ its vector potential. Since in the recollision classical model, the electron-hole trajectories are computed via integrating the group velocity, we have that the group velocity reads:
\begin{eqnarray}
v_e(t)&=&\frac{\partial \epsilon_{\mathrm c}(q)}{\hbar \partial q} = v_{\mathrm c} \sin q(t) = v_{\mathrm c} \sin [A(t) + q(0)],\nonumber\\
v_h(t)&=&\frac{\partial \epsilon_{\mathrm v}(q)}{\hbar \partial q} = v_{\mathrm v} \sin q(t) = v_{\mathrm v} \sin [A(t) + q(0)].
\end{eqnarray}



Using the velocity, we obtain the position of electron and hole:
\begin{eqnarray}
x_e(t)&=&\int_0^t v_e(t')dt',\nonumber\\
x_h(t)&=&\int_0^t v_h(t')dt'.
\end{eqnarray}

At the recollison time $t_c$, the electron and hole return to the same position in spatial space, that is~\cite{Vampa2015,Vampa2017}
\begin{equation}
x_e(t_c)=x_h(t_c).
\end{equation}

Once we get the solution of $t_c$, we will have the energy difference between electron and hole at time $t_c$:
\begin{equation}
\Delta E(t_c)= \epsilon_{\mathrm c}(q(t_c)) - \epsilon_{\mathrm v}(q(t_c)).
\end{equation}
This energy difference will transfer to phonon emission.

In Fig.~\ref{sfig:trajectory}, we show the semi-classical solution of electron and hole trajectory,
and  the harmonic photon energy as a function of the recollision time.
One can see that the recollision almost occurs within one optical cycle.
Although the analysis here is semi-classical, 
this observation is largely consistent with the Gabor time-profile analysis in the main text (Fig. 3).
This calculation provides a way to understand the trajectories observed in HHG spectra.

\begin{figure}[!htb]
	\includegraphics[width=0.5\textwidth]{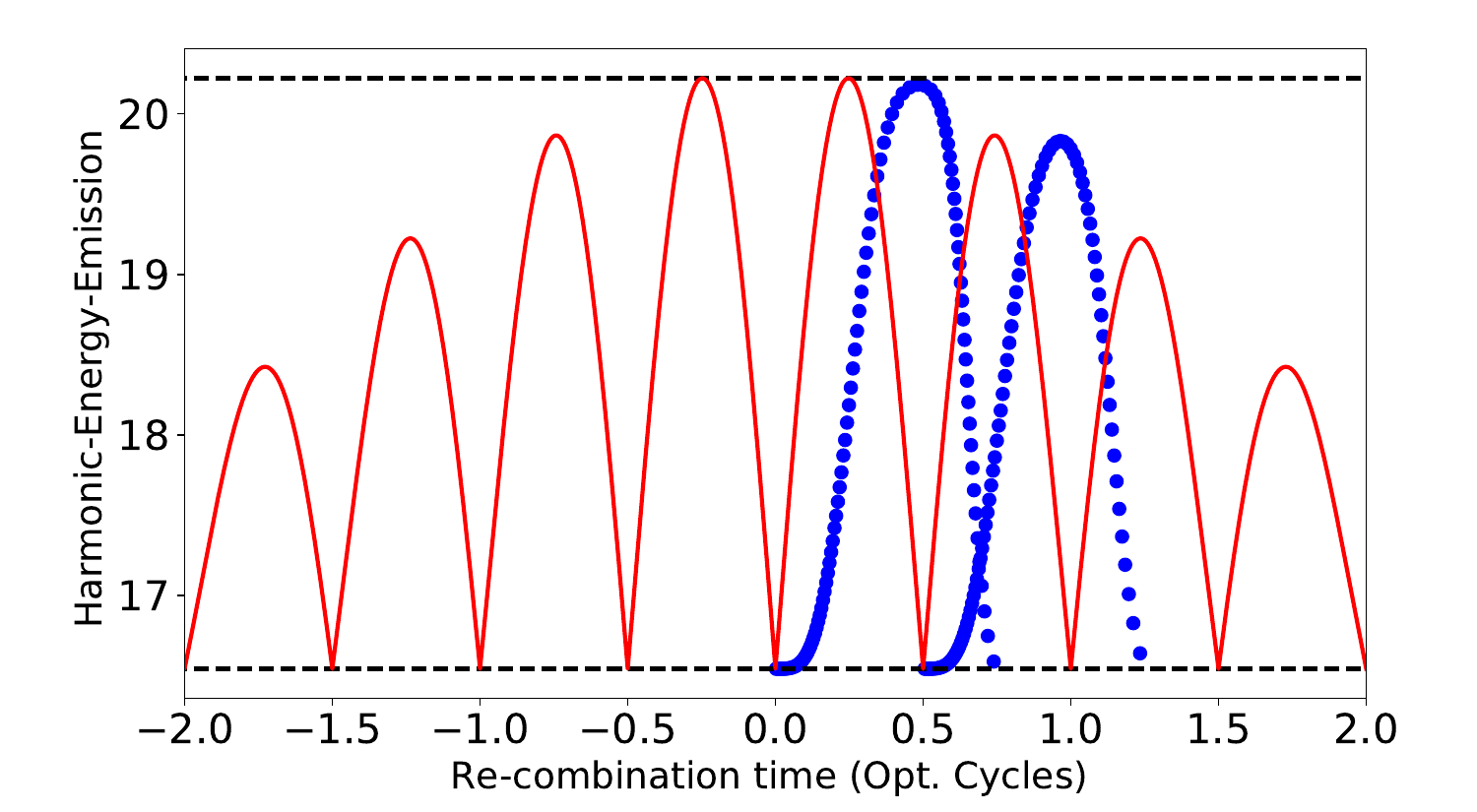}
	\includegraphics[width=0.51\textwidth]{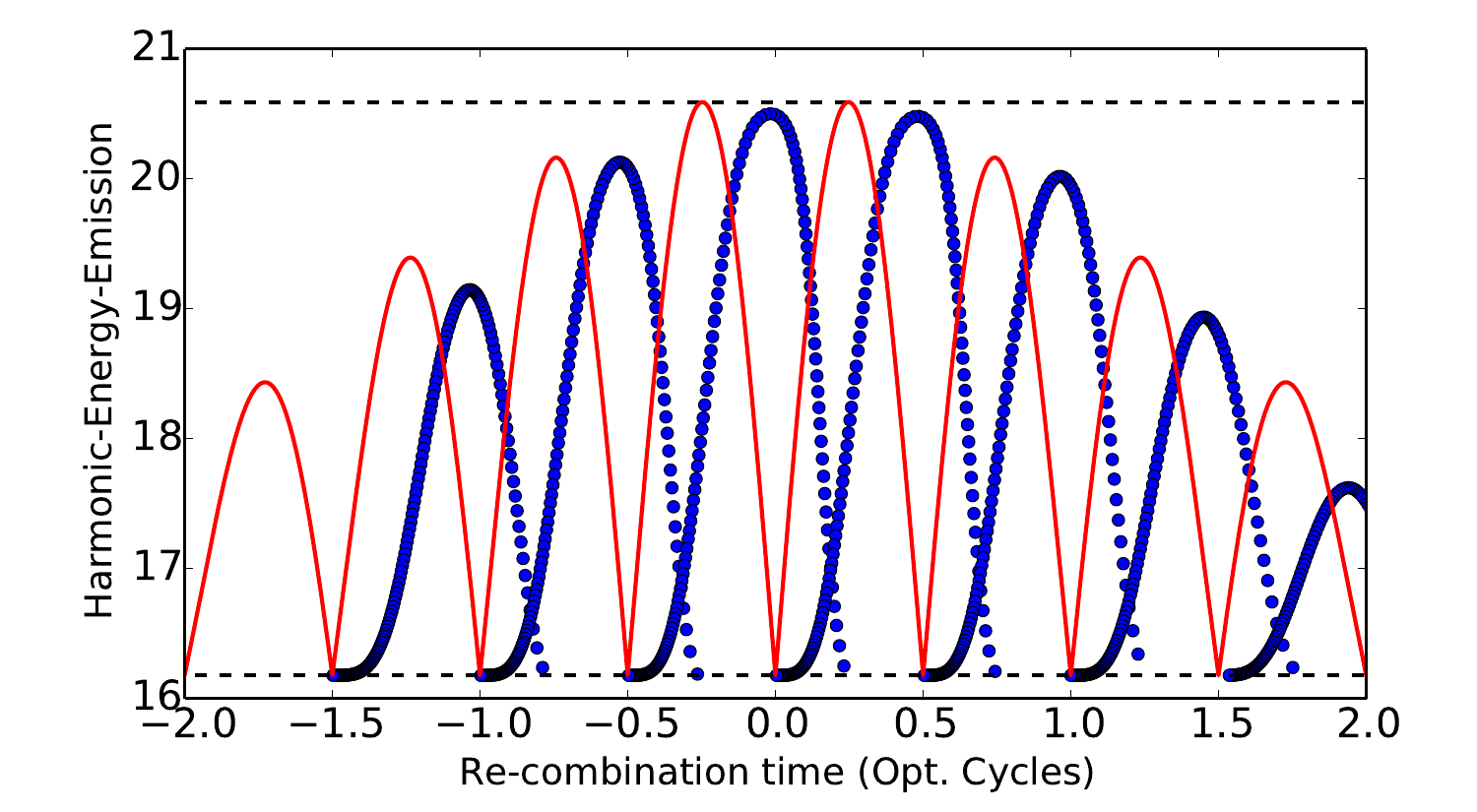}
	\caption{(Top and Left) Dispersion relation of conduction band  and valence band. (Top and Right) Trajectory of electron (blue) and hole (green) as a function of time.
		(Bottom) The resulting phonon emission as a function of the recollision time $t_c$ (blue dots). The red line shows the derivative of electric field.
	} \label{sfig:trajectory}
\end{figure}

\section{Mean Field Solution of Kondo Breakdown}

In this section, we will explore the mean field solution of the time-dependent Kondo lattice model:
\begin{equation}\label{eq:ham_mf}
\hat H=  \sum_{ i,\sigma}^{L-1}[ J_0 e^{-i\Phi(t)} c^{\dagger}_{i\sigma} c_{i+1,\sigma} +h.c.] + J_K \sum_i \mathbf S_i \cdot \mathbf s_i.
\end{equation}
$c^\dagger_{i,\sigma}$ denotes the creation operator of a conduction electron in the state with spin $\sigma$. The $\mathbf S_i$ is localized moment with spin-$1/2$.
And each localized moment interacts via an exchange coupling $J_K$ with spin of conduction electron, where spin degree of freedom of conduction electron
is defined as $\mathbf{s}_i=\frac{1}{2}\sum_{\alpha,\beta}c^\dagger_{i\alpha} \sigma_{\alpha\beta}c_{i\beta}$. Next we set $4J_0=1$ in the calculations.

We first decouple the local moment using fermionic operators
\begin{equation}
S^z_i=\frac{1}{2}(f^\dagger_{i\uparrow} f_{i\uparrow} - f^\dagger_{i\downarrow}f_{i\downarrow}), S^\dagger_i=f^\dagger_{i\uparrow} f_{i\downarrow}, S^-=f^\dagger_{i\downarrow} f_{i\uparrow}
\end{equation}
and $f^\dagger_{i\uparrow}f_{i\uparrow}+f^\dagger_{i\downarrow}f_{i\downarrow}=1$ should be satisfied. Thus, the Kondo coupling term becomes
\begin{equation}
J_K \sum_i \mathbf{S}_i \cdot \mathbf{s}_i = J_K\sum_i \frac{1}{2}[f^{\dagger}_{i\uparrow}f_{i\downarrow}c^{\dagger}_{i\downarrow}c_{i\uparrow}+
c^{\dagger}_{i\uparrow}c_{i\downarrow}f^{\dagger}_{i\downarrow}f_{i\uparrow} ] +
\frac{1}{4} [f^\dagger_{i,\uparrow} f_{i,\uparrow}-f^\dagger_{i\downarrow}f_{i\downarrow}][c^\dagger_{i\uparrow}c_{i\uparrow}-c^\dagger_{i\downarrow}c_{i\downarrow}]
\end{equation}

In the mean-field approximation, we have
\begin{eqnarray}
f^{\dagger}_{i\uparrow}f_{i\downarrow}c^{\dagger}_{i\downarrow}c_{i\uparrow}&=&
\langle f^{\dagger}_{i\uparrow} c_{i\uparrow}\rangle f_{i\downarrow}c^{\dagger}_{i\downarrow} +
f^{\dagger}_{i\uparrow} c_{i\uparrow} \langle f_{i\downarrow}c^{\dagger}_{i\downarrow}\rangle -
\langle f^{\dagger}_{i\uparrow} c_{i\uparrow}\rangle \langle f_{i\downarrow}c^{\dagger}_{i\downarrow}\rangle \\
c^{\dagger}_{i\uparrow}c_{i\downarrow}f^{\dagger}_{i\downarrow}f_{i\uparrow}&=&
\langle c^{\dagger}_{i\uparrow} f_{i\uparrow}\rangle c_{i\downarrow}f^{\dagger}_{i\downarrow} +
c^{\dagger}_{i\uparrow} f_{i\uparrow} \langle c_{i\downarrow}f^{\dagger}_{i\downarrow}\rangle -
\langle c^{\dagger}_{i\uparrow} f_{i\uparrow}\rangle  \langle c_{i\downarrow}f^{\dagger}_{i\downarrow}\rangle \\
f^{\dagger}_{i\uparrow}f_{i\uparrow}c^{\dagger}_{i\uparrow}c_{i\uparrow}&=&
f^{\dagger}_{i\uparrow}f_{i\uparrow} (1-c_{i\uparrow}c^\dagger_{i\uparrow})=
f^{\dagger}_{i\uparrow}f_{i\uparrow} + \langle f^{\dagger}_{i\uparrow} c_{i\uparrow}\rangle f_{i\uparrow} c^\dagger_{i\uparrow} +
f^{\dagger}_{i\uparrow} c_{i\uparrow} \langle f_{i\uparrow} c^\dagger_{i\uparrow}\rangle - \langle f^{\dagger}_{i\uparrow} c_{i\uparrow}\rangle \langle f_{i\uparrow} c^\dagger_{i\uparrow}\rangle
\end{eqnarray}
and
\begin{eqnarray}
(n^f_{i\uparrow}-n^f_{i\downarrow}) (n^c_{i\uparrow}-n^c_{i\downarrow}) &=&
\langle n^f_{i\uparrow}-n^f_{i\downarrow}\rangle (n^c_{i\uparrow}-n^c_{i\downarrow}) +
(n^f_{i\uparrow}-n^f_{i\downarrow}) \langle n^c_{i\uparrow}-n^c_{i\downarrow}\rangle -
\langle n^f_{i\uparrow}-n^f_{i\downarrow}\rangle \langle n^c_{i\uparrow}-n^c_{i\downarrow}\rangle
\end{eqnarray}

The hamiltonian becomes
\begin{eqnarray}
H_{MF}&=&J_0\sum_{i,\sigma} c^\dagger_{i,\sigma} c_{i+1,\sigma} e^{-i\Phi(t)} + h.c \\
&&+ J_K  \frac{1}{2} \sum_i h_{i,\uparrow}f_{i\downarrow}c^{\dagger}_{i\downarrow} + h^*_{i,\downarrow} c_{i\uparrow}f^{\dagger}_{i\uparrow} + h^*_{i,\uparrow} c_{i\downarrow}f^\dagger_{i\downarrow} + h_{i,\downarrow}f_{i\uparrow}c^\dagger_{i\uparrow}
+ h_{i\uparrow}h^*_{i\downarrow} +h^*_{i\uparrow} h_{i\downarrow} \\
&&+ J_K  \frac{(1-x)}{4} \sum_i f^\dagger_{i\uparrow}f_{i\uparrow}+ h_{i,\uparrow}f_{i\uparrow}c^{\dagger}_{i\uparrow} + h^*_{i,\uparrow}c_{i\uparrow}f^{\dagger}_{i\uparrow}  +
f^\dagger_{i\downarrow}f_{i\downarrow}+ h_{i,\downarrow}f_{i\downarrow}c^{\dagger}_{i\downarrow} + h^*_{i,\downarrow}c_{i\downarrow}f^{\dagger}_{i\downarrow}
+ h_{i\uparrow}h^*_{i\uparrow} +h^*_{i\downarrow} h_{i\downarrow} \\
&&+J_K \frac{x}{4} \sum_i \overline{S^z_{i}} n^c_{i,\uparrow} - n^c_{i\downarrow} \overline{S^z_{i}} + n^f_{i,\uparrow} \overline{s^z_{i}}- \overline{s^z_{i}} n^f_{i,\downarrow} - \overline{S^z_{i}} \overline{s^z_{i}}
\end{eqnarray}
where the mean-field parameters are
\begin{equation}
\overline{S^z_i}=\langle n^f_{i\uparrow}-n^f_{i,\downarrow}\rangle, \overline{s^z_i}=\langle n^c_{i\uparrow}-n^c_{i,\downarrow}\rangle, h_{i,\uparrow}= \langle f^{\dagger}_{i\uparrow} c_{i\uparrow}\rangle, h_{i,\downarrow}= \langle f^\dagger_{i\downarrow} c_{i\downarrow}\rangle,
\end{equation}
and $h_{i,\uparrow}, h_{i,\downarrow}$ describe the hybridization between c-electron and f-electron, which can be taken as an order parameter for Kondo insulator.
$x$ is a decoupling parameter controlling the weight of different mean-field channels.~\cite{Senthil2004} Here we choose $x=0.2$ in our calculation.  
That is, in the density-density interaction $S^z_i s^z_i$, the weight in mean-field channel Eq. (C10) is $0.8$ and in the channel of Eq. (C11) is $0.2$.

At initial time $t=0$, we first get the mean field solution self-consistently. For $J_K\neq0$, the mean field solution are: $\overline{S^z_i}=0,\overline{s^z_i}=0,  h_{i,\uparrow}= h_{i,\downarrow}\neq0$, which relates to the Kondo insulator phase with non-zero hybridization between local moments and conduction electrons.

For time evolution at $t\neq0$,  the dynamics of the system are determined by the following time-dependent equation
\begin{equation*}
i\frac{d}{dt} \left(
\begin{array}{c}
c^\dagger_{\{i\}} \\
f^\dagger_{\{i\}} \\
\end{array}
\right) = H_{MF}(h_{\{i\},\uparrow}(t),h_{\{i\},\downarrow}(t))
\left(
\begin{array}{c}
c^\dagger_{\{i\}}(t) \\
f^\dagger_{\{i\}}(t) \\
\end{array}
\right).
\end{equation*}
Here, $H_{MF}$ is the time-dependent Hamiltonian in which the hybridization parameters $h_{\{i\},\uparrow}(t),h_{\{i\},\downarrow}(t)$ now evolves in time.

We show the mean field phase diagram is shown in Fig. \ref{sfig:mf}.
The Kondo insulator phase is identified by hybridization parameters $ h_{\{i\} \uparrow(\downarrow)}(t)\neq 0$. The transient transition point (open square) is determined by hybridization parameters becomes zero: $ h_{\{i\},\uparrow(\downarrow)}(t_c) =0$.
In Fig. \ref{sfig:mf} (left), the laser field can drive a transition from the Kondo insulator to a liquid phase when $J_K$ is below a critical value ($J_K<J_K^c$). When $J_K$ is strong enough ($J_K>J_K^c$), the Kondo insulator can be not destroyed.
This mean field phase diagram captures the main features shown in Fig. 1(c-d) in the main text.
Moreover, we also study the two-dimensional Kondo lattice model under the driven laser field, as shown in Fig. \ref{sfig:mf} (right).
We find very similar feature as in the one-dimensional case: The driven laser field can destroy the Kondo insulator phase when $J_K$ is smaller than a critical value,
while Kondo insulator phase is robust when $J_K$ is strong enough. 
This mean-field calculation indicates that the Kondo collapse is general for driven Kondo systems,
despite that our full quantum mechanical method is only limited to one-dimension.

\begin{figure}[b]
	\includegraphics[width=0.25\textwidth]{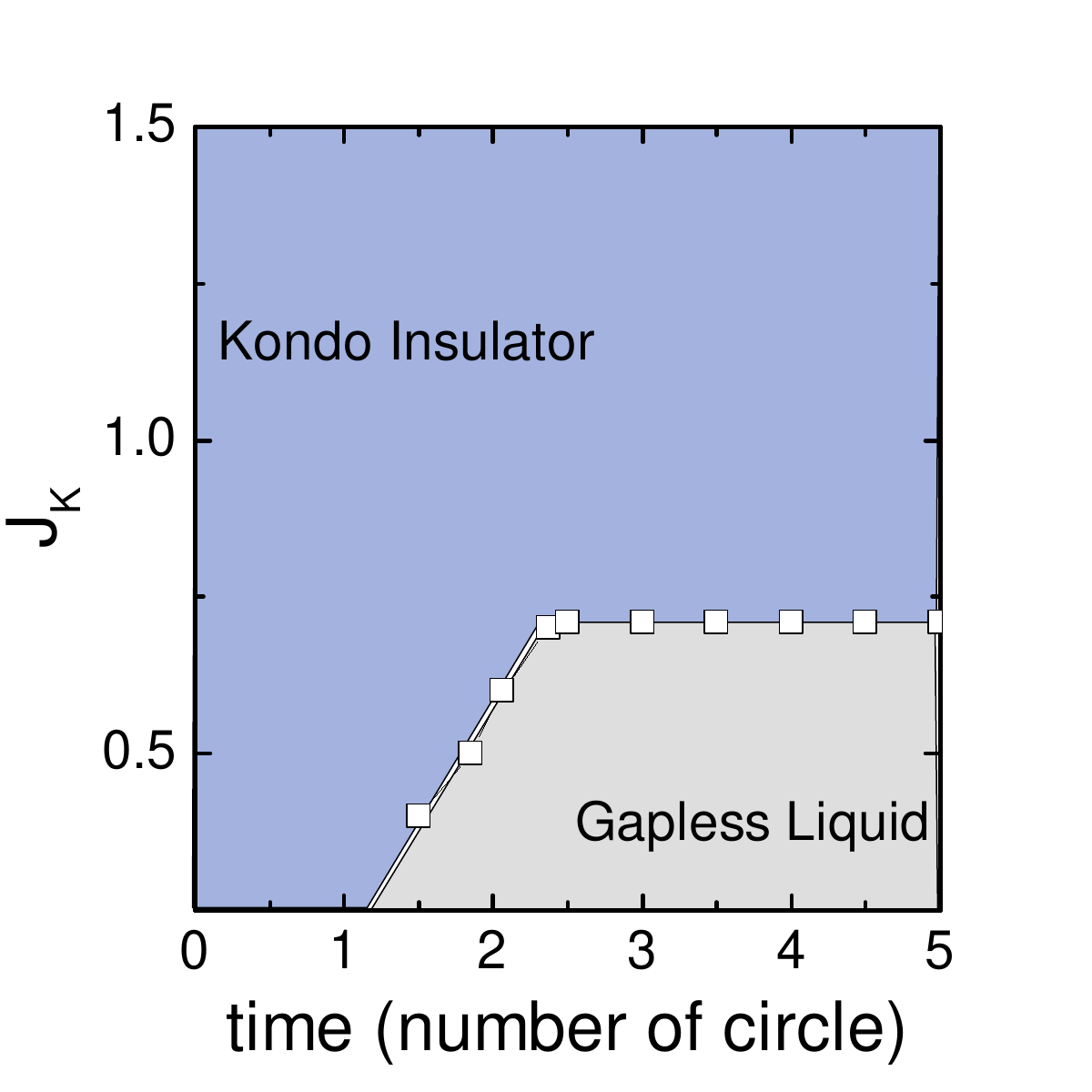}
	\includegraphics[width=0.25\textwidth]{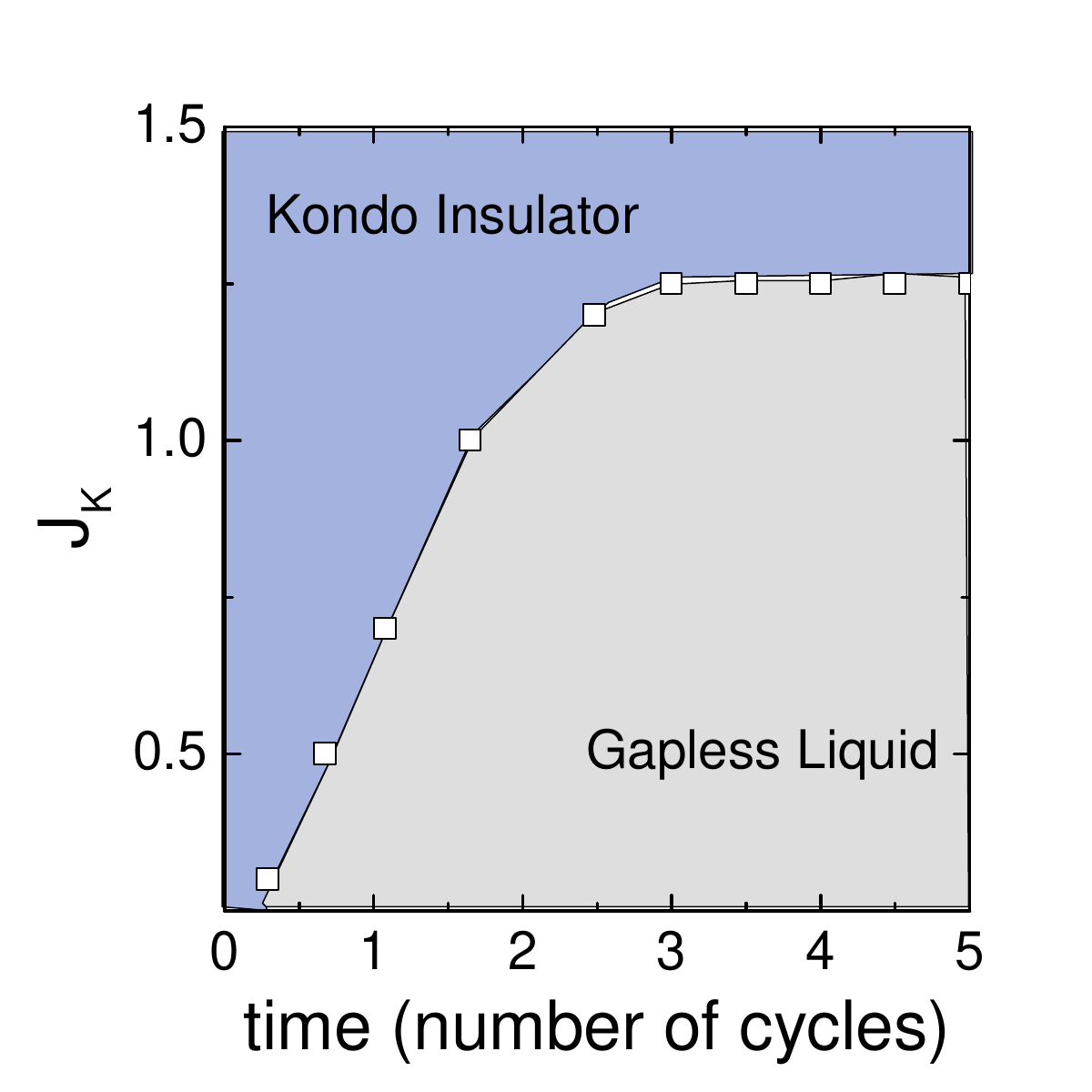}
	\caption{(Left) The mean-field phase diagram of time-dependent Kondo lattice model (Eq. (12)). 
		(Right) The mean-field phase diagram of two-dimension Kondo lattice model. We choose the square lattice as an example.
		The Kondo insulator phase is identified by hybridization parameters $h_i(t)\neq 0$, and the transition time $t_c$ is identified by $h_i(t_c) =0$.  The laser field has peak strength $E_0 = 0.16$ a.u., frequency $\omega_0 = 0.02$ a.u. and a 5-cycle $\sin^2$ envelope.
	} \label{sfig:mf}
\end{figure}

%
%
\end{widetext}

\end{document}